# Electromagnetism in the Encyclopaedias

Isobel Falconer

This article is based on a talk given for the IoP History of Physics Group meeting to celebrate the 200$^{th}$ anniversary of Ørsted's discovery, 16 Sept 2020

"Heat, Electricity and Magnetism… though very different, agree in some general characters" wrote the natural philosopher John Playfair in 1819 in his influential "Dissertation Second" for the *Encyclopaedia Britannica*. His essay provides little evidence to support the claim of John Leslie, writing a sequel dissertation in the 1820s, that, "[Electricity's] close connexion, if not identity, with magnetism, had been long suspected."  Leslie, of course, was writing with the benefit of hindsight, informed by Hans Christian Ørsted's discovery of the effect of an electric current on a magnetised needle in 1820. After a brief overview of Ørsted's life, I investigate what encyclopaedia articles such as these can tell us about electricity and magnetism in the years leading up to Ørsted's publication, and immediately afterwards.

## Oersted's Life[1]
Hans Christian Ørsted was born on 14 March 1777 in Rudkøbing, Island of Langeland, Denmark (Figure 1), where his parents, Søren Christian Ørsted and Karen (nee Hermansdatter), ran a pharmacy. His brother, Anders Sandøe, was born only 20 months later; he eventually became a lawyer, member of the Danish government, and (briefly) Prime Minister. The two boys, who were viewed as gifted, were educated together in a sort of community endeavour. They learnt chemistry and law from a Norwegian student and a district judge, respectively, and mathematics, French and English from a land surveyor in return for accommodation at the pharmacy guest house. They also had free access to the private libraries of the better off local citizens.

In 1794 Hans and Anders proceeded, together, to the University of Copenhagen, where Anders read law and Hans studied medicine, focusing on chemistry and pharmacy. In 1797 he graduated with a degree in pharmacy. Both brothers attended Børge Riisbrigh's lectures on the philosophy of Immanuel Kant, which was still very new (his *Metaphysical Foundations* was published in 1786). Ørsted had already been shocked by the conflicting opinions, and lack of evidence, for the composition of various bodily fluids, and turned to Kant's ideas as the way out of the mess. His 1799 PhD dissertation, *Structure of the Elementary Metaphysics of External Nature*, attempted to show how adherence to Kant's principles could provide a more robust and better evidenced basis for chemistry. According to Jackson and Jelveds, Ørsted interpreted these principles as:
1. that "true" science must be based on a number of "necessary" facts of unquestioned validity,
2. that it must be possible to deal with these facts mathematically.[2]

---

[1] This section is heavily based on the useful article by Andrew D. Jackson and Karen Jelveds, "H. C. Ørsted and the Discovery of Electromagnetism" (2020)
https://www.royalacademy.dk/~/media/RoyalAcademy/Filer/Artikler/HCOEM.pdf [accessed 8 Feb. 2021]
[2] Jackson & Jelveds, p.5

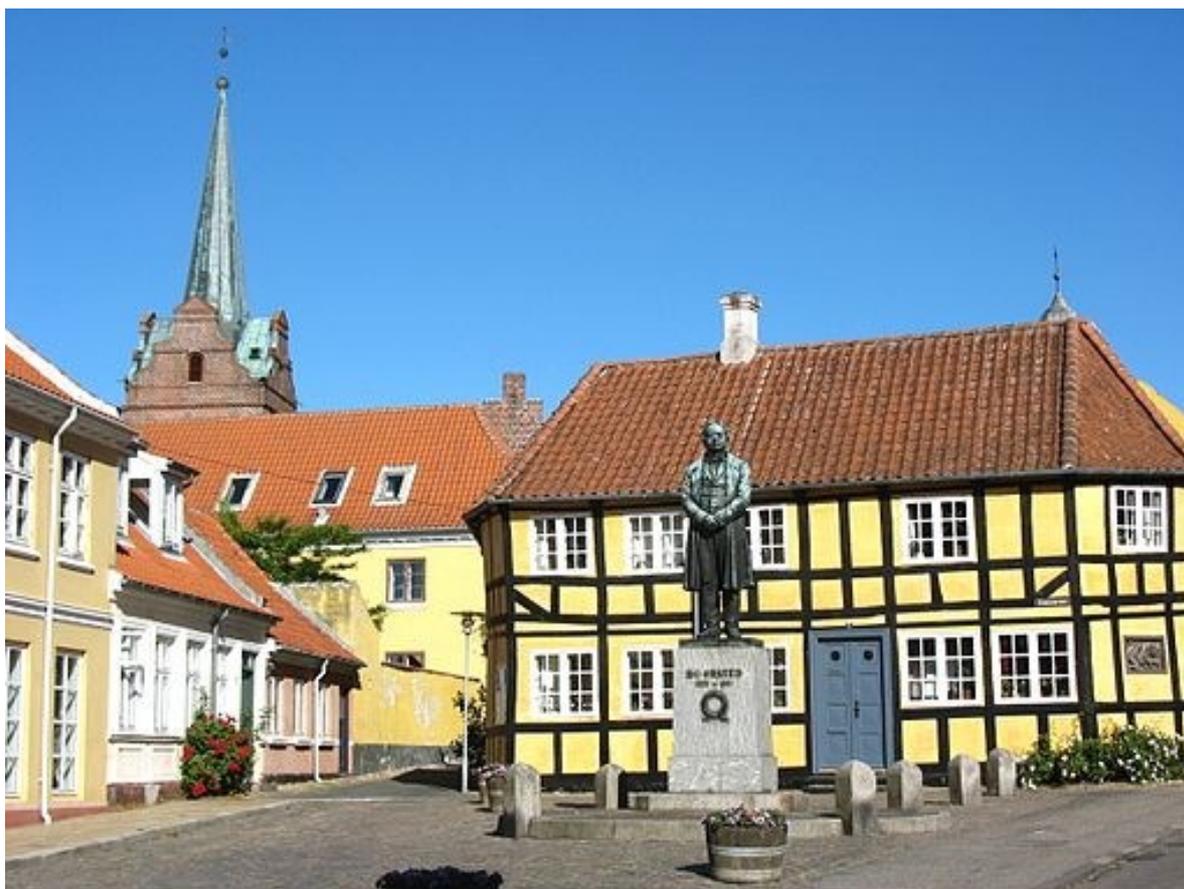

Figure 1. Photo of Rudkøbing, with Ørsted's statue in the square. Photograph by Hubertus45 - Own work, CC BY-SA 3.0, https://commons.wikimedia.org/w/index.php?curid=4396729

Kant discussed his ideas in relation to physics, but Ørsted extended them to chemistry based on two further claims Kant had made about the constitution of matter. First that there are fundamental *repulsive* and *attractive* forces which both prevent matter from collapsing and hold it together. Second, that matter is infinitely divisible, leading to the conclusion that atoms do not exist. Taking these claims and the first principle together, Ørsted pursued a search for the observable consequences of infinitely divisible matter and of the interactions between forces of nature. He believed that the many observable forces resulted from different mixtures of the two fundamental forces, seeking this underlying unity of forces. Quotations from 1803 and 1821 demonstrate the persistence of his belief:

> *The constituent principles of heat, which are important in alkalis and acids, in electricity, and in light, are also the principles of magnetism, and thus we would have the unity of all the forces which act together to govern the entire universe, . . . for do friction and impact not produce both heat and electricity, and are dynamics and mechanics not thereby perfectly intertwined?*[3]
>
> *A long time ago the author himself adopted a system according to which all internal effects in bodies, such as electricity, heat, light, as well as chemical combinations and dissociations are due to the same fundamental forces. This system, which he has advanced in a few earlier treatises, has been developed more completely in his*

---

[3] Hans Christian Ørsted (1803) *Materials for a Chemistry of the Nineteenth Century*, trans. in K. Jelveds, A.D. Jackson & O. Knudsen (eds), *Selected Scientific Works of Hans Christian Ørsted,* (Princeton University Press, 2014) pp.120-165

> Ansichten [sic] der chemischen Naturgesetze*, which was published in 1812, and already then he arrived at the result that magnetism must be produced by electrical forces in their most bound form.*[4]

Following his PhD, Ørsted obtained a bursary to tour France and Germany and study technical chemistry – porcelain manufacture and brewing. In Germany he met William Ritter, another enthusiastic disciple of Kant, and a promising physicist who constructed the first dry cell in 1803.[5] Together they studied the works of a third Kant enthusiast, the Hungarian chemist Jacob Joseph Winterl, who claimed to have discovered the two physical substances, "andronia" and "thelyke", which were supposedly to be the embodiment of acidity and alkalinity, respectively. Ørsted's uncritical support of Winterl's ideas and misguided championship of some of Ritter's less verifiable results, undermined his credibility. After his return to Copenhagen in 1804 he was appointed a "professor extraordinarius" (lecturer), being deemed too interested in philosophy to hold a full professorship.

In an attempt to provide further support for his Kantian views, Ørsted devoted a lot of time to studying acoustic figures, using a Chladni plate. If, as he supposed, mechanical and electric forces were related, the mechanical oscillations of the plate would produce static electrical effects. He used fine lycopodium powder on the plate instead of the much more usual, but coarse, sand. He reasoned that lycopodium would stick to the positively charged areas of the plate, and that on the rest would easily be knocked off. He was successful in producing such figures, and in printing them by pressing paper smeared with gum arabic up against the plate - a forerunner of the xerox machine! He published *Experiments on Acoustic Figures* in 1810. In 1812 Ørsted drew all his work so far together in his *View of the chemical laws of nature obtained through recent discoveries*, the work in which he first suggested explicitly that magnetism was produced by electric forces.

By 1814, Ørsted was sufficiently established to marry Inger Birgitte Ballum. The couple had seven children. In 1817 he was finally appointed a full professor at the University of Copenhagen. In the same year he began a study of the compressibility of gases that he pursued on and off for much of the rest of his life. The study began in an attempt to disprove Newton's atomic theory by showing that the perfect gas laws held at all pressures, i.e. there was no characteristic size of atoms beyond which a gas could not be further compressed. Initial experiments upheld his hypothesis; when in 1825 he found that sulphur dioxide deviated from the law near phase transitions, he quietly dropped the implications for atomic theory, while holding that the laws were obeyed except in these narrow instances.

By this time, though, he was world famous for his 1820 demonstration that an electric current exerted a twisting force on a magnetic needle.[6] The experiment had an impact similar to that of Röntgen's discovery of x-rays 75 years later. In Britain, Humphry Davy rapidly reproduced the effect and secured the Copley medal for him. In France Arago demonstrated the effect to the French Academie Royale on 11 September, and two weeks later Ampère

---

[4] Hans Christian Ørsted (1821) "Note on the Discovery of Electromagnetism" trans. in Jelveds, Jackson & Knudsen, *Selected Scientific Works,* pp.425-429

[5] Richter, K. (2016). *Das Leben des Physikers Johann Wilhelm Ritter: Ein Schicksal in der Zeit der Romantik*. Springer-Verlag.

[6] Ørsted's ensured that a translation of his account of the experiment appeared in "Experiments on the Effect of the Electric Conflict on the Magnetic Needle", *Annals of Philosophy*, Vol. 16 (1820), pp. 273-76. The volume is available via the Hathi Trust at https://babel.hathitrust.org/cgi/pt?id=osu.32435051156651 [accessed 8 Feb 2021]

demonstrated the corresponding interaction between two current-carrying wires.[7] Ørsted was made a Member of the Prussian Academy of Sciences in 1820, a Fellow of the Royal Society in 1821, a Foreign Member of the Swedish Academy of Sciences in 1822, and a Member of the Academie Royale in 1823.

Ørsted visited Germany, France, and Britain between 1820 and 1823. However, apart from his 1823 work with Joseph Fourier on the thermo-electric pile, he did not pursue electromagnetism much further. Instead, he resumed his work on the gas laws, and was the first to isolate aluminium in 1825. Inspired by what he had seen of the scientific infrastructure in France and Britain - the École Polytechnique, and the many scientific societies springing up in London - he turned much of his attention to fostering similar developments in Denmark. In 1824 he founded the Danish Society for the Propagation of Natural Sciences, modelled on London's Royal Institution, and in 1829 he took over the lead of plans for "Den Polytekniske Læreanstalt" (College of Advanced Technology, later the Technical University of Denmark), insisting that the College be scientifically as well as practically oriented. He became its first Director, a position that he occupied until his death in 1851.[8]

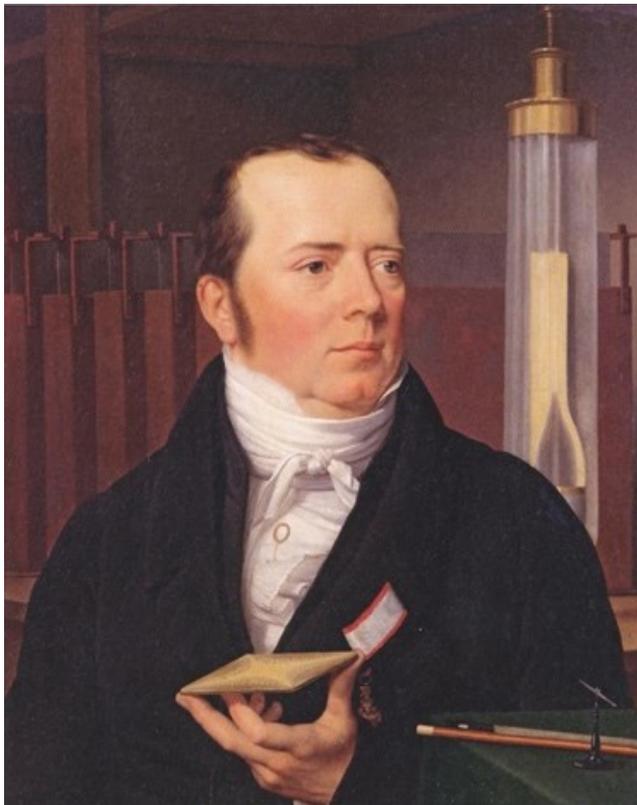

Figure 2. Painting of Ørsted in 1823 by Christoffer Wilhelm Eckersberg. Ørsted is holding his Chladni plate, while a compass needle is less prominent on the table on the right. A very large voltaic pile is behind him on the left and his piezometer back right. Placed in the Public Domain by the Danmarks Tekniske Museum,
https://commons.wikimedia.org/w/index.php?curid=387999

---

[7] For a full account of the immediate uptake and consequences of Ørsted's discovery see Friedrich Steinle, trans. Alex Levine (2016). *Exploratory Experiments: Ampère, Faraday, and the Origins of Electrodynamics,* University of Pittsburgh Press

[8] Niels Bohr Institute (2010) ' Inspiration from Europe - plans in Copenhagen'
https://www.nbi.ku.dk/english/www/hco/oersted/uddannelsen/  [accessed 8 Feb 2020]

## The Encyclopaedia Britannica

Investigation of *Encyclopaedia Britannica* (*EB*) articles around 1820 gives us an excellent view of understandings of electricity and magnetism at the time.[9] The Edinburgh-based *Encyclopaedia* went through a number of editions and supplements between 1800 and 1830. Each edition appeared in numerous instalments over several years, the instalments being later collected into volumes. Thus, initial writing and distribution of an article was sometimes up to ten years prior to the date on the bound volume in which it appeared. The 4th edition was completed in 1810 and the publisher, Constable, began planning a Supplement, to update those articles where knowledge was changing fast. But, at the same time, he reprinted the 4th edition with minor changes, as the 5th and 6th editions, to maintain his income flow. Thus, the Supplement, with its recent updates became the Supplements to the 4th, 5th, and 6th editions, as shown in Table 1. Each volume of the Supplement was prefaced by a lengthy "dissertation" giving a history of progress in a particular subject. Further "dissertations" subsequently accompanied some volumes of the 7th (and 8th) editions.

- 1797 3$^{rd}$ edn. "Electricity", "Magnetism" by **James Tytler** (Edinburgh apothecary and editor of the *EB*)
- 1800 Supplement to 3$^{rd}$ edn. "Electricity" "Magnetism" by **John Robison** (Professor of Natural Philosophy, Edinburgh)
- 1804-1810 4$^{th}$ edn. "Electricity", "Magnetism" by **Jeremiah Kirby** (Edinburgh doctor); "Galvanism" by **James Millar** (medical doctor and editor of the *EB*)
- 1810-1823 5$^{th}$ & 6$^{th}$ edns reprinted 4$^{th}$ edn
- 1813-1824 Supplement to 4$^{th}$, 5$^{th}$ & 6$^{th}$ edns. "Dissertation Second Exhibiting a General View Of the Progress of Mathematical and Physical Science since the Revival of Letters in Europe" by **John Playfair** (Professor of Natural Philosophy, Edinburgh); "Electricity", "Galvanism" by **Jean Baptiste Biot** (French Academy of Sciences)
- 1830-1842 7$^{th}$ edn. "Dissertation Fourth; Exhibiting a General View of the Progress of Physical Science Chiefly During the Eighteenth Century" by **John Leslie** (Professor of Natural Philosophy, Edinburgh); "Electricity", "Magnetism", "Voltaic Electricity" by **David Brewster** (Edinburgh scientist and journal editor)

Table 1. Editions of the *Encyclopaedia Britannica* (*EB*) and authors of articles on electricity, magnetism, etc.

Sir John Leslie, Professor of Natural Philosophy at Edinburgh, was the first person after Ørsted's discovery to write about electromagnetism. He did so in his "Dissertation Fourth on the Progress of Mathematical and Physical Science", which must have been written at some time between 1827, when commissioning for the 7th edition began, and 1832 when he died. Leslie tells us that he originally intended to finish his account of science at around 1800, but that Ørsted's discovery was so outstanding that he included it:

> *The original design of this discourse was to come no lower than the early part of the present century, and to avoid discussing the merits of contemporaries. But I cannot resist the pleasure of noticing the signal advance which Electricity has lately made. Its close connexion, if not perfect identity, with Magnetism had been long suspected, and was even adopted by several ingenious theorists.... Professor Oerstedt [sic] of*

---

[9] Digitised copies of all these editions are available from the National Library of Scotland
https://digital.nls.uk/encyclopaedia-britannica/archive/188936619

> *Copenhagen… published near the close of the year 1820 his great discovery, which awakened the public attention, and gave rise to numerous speculations that frolic in the giddy maze of electric and magnetic currents.*[10]

Leslie is emphatic about the unifying outcome of Ørsted's work – confirmed evidence of the "close connexion" of electricity and magnetism – but his understanding of this connection is far from our modern one. He continued:

> *It must indeed be confessed, that after all the progress which Electricity and its younger branch Galvanism have made, the hypotheses commonly received are exceedingly vague and unphilosophical... It is rather amusing to observe the complacency with which some ingenious persons describe the play and vagaries of an Electrical Current, whose existence was never proved. We are acquainted only with electric attraction and repulsion, and with the transmission of electric influence: All beyond these elementary principles, rests on hasty conjecture. Instead of adopting one or two fluids, it were safer to suspend the assumption of any….* [emphasis added][11]

Notably, Leslie makes no mention of Ampère who, in subsequent historical accounts has assumed a much greater importance than Ørsted. Ampère was later lauded for having developed a mathematical theory of electromagnetism, which Ørsted did not. However, Leslie, who had previously been a Professor of Mathematics and was in a position to appreciate mathematised theory, possibly dismisses Ampère as one of the frolickers in the "giddy maze". His is the account of one writing before the strong tradition of mathematical electromagnetism instituted by William Thomson and James Clerk Maxwell took effect, colouring later perceptions.[12]

Leslie's account raises two issues for further investigation:
1. Leslie's claim that electricity and magnetism had been long suspected to be similar
2. The ways in which current electricity or galvanism was understood in the early 19th century

## Changing views on the connection between electricity and magnetism

Leslie's claim about the longevity of the connection between electricity and magnetism can be traced back through editions of the *Encyclopaedia.* It becomes apparent that although Leslie is correct, he fails to mention that many other phenomena had also been promoted as intimately connected with magnetism, often with no specific preference assigned to electricity.

In "Magnetism" in the 3rd edition, published around 1797, James Tytler (the probable author) was swayed by the finding that an iron bar struck by lightning or an electric spark became magnetised. He concluded that "… the analogies betwixt magnetism and electricity are so great, that the hypothesis of a magnetic as well as of an electric fluid has now gained general credit."[13] Note, though, that these were two *different* fluids; despite their similarity he contrasted electricity, in which the "fluid" is often "perceptible by our senses", with magnetism whose cause could never be rendered "perceptible otherwise than by its effects."[14] This was a major methodological problem at a time when sensory experience was intrinsic to scientific experimentation, and it crops up repeatedly in other accounts also.[15]

---

[10] John Leslie (1842), "Dissertation Fourth; Exhibiting a General View of the Progress of Physical Science Chiefly During the Eighteenth Century", *Encyclopaedia Britannica,* 7th edn, vol.1 (Edinburgh, A&C Black) pp. 573-677, on p.623.
[11] Leslie, "Dissertation Fourth", p.624
[12] For an account of one way in which Thomson and Maxwell wrote history to bolster a mathematical approach to electromagnetism see Isobel Falconer (2017), "No actual measurement … was required: Maxwell and Cavendish's null method for the inverse square law of electrostatics", *Studies in History and Philosophy of Science Part A*, *65–66*, 74–86. https://doi.org/10.1016/j.shpsa.2017.05.001

The corresponding article for electricity contained sections on "Proofs of the identity of the electric fluid and the elementary fire or light of the sun", "Effects of electricity on vegetation", and "Effects of electricity on animals." Tytler drew multiple connections, not just with magnetism:
> *We shall now find [electricity] guiding the planets in their courses… giving stability and cohesion not only to terrestrial substances, but to the globe of earth itself…. For a further account of the operations of this fluid in producing the phenomena of nature, see the articles Atmosphere, Aurora Borealis, Earthquake, Hail, Hurricane, Lightning, Meteor, Rain, Snow, &c.*[16]

John Robison, Professor of Natural Philosophy at Edinburgh, updated the articles on "Electricity" and "Magnetism" around 1800 for the Supplement to the 3rd edition. His was a strongly Newtonian approach, which favoured mathematical theories based on attractive and repulsive forces and downplayed the importance of phenomenological associations. He discounted the relevance of the evidence that sparks can magnetise needles on the grounds that "the polarity that [the spark] gives is always the same with that given by great heat; and there is always heat in this operation."[17] After a careful comparison of electric and magnetic phenomena, and the explanatory power of their various theories, Robison concluded "that the electric and magnetic fluid are totally different," although they might "be related by means of some powers hitherto unknown."[18]

Robison's lead was followed by Jeremiah Kirby, the Edinburgh doctor who wrote "Electricity" and "Magnetism" for the 4th edition of the *Encyclopaedia,* published in 1804 to 1810. Kirby considered that "there are several circumstances which show their [electricity and magnetism's] original causes to be different." Electricity, though, he considered closely related to both caloric (the substance of heat) and light, indeed it might be a compound of them:
> *What has been now said is, we think, sufficient to prove, that the electric fluid is neither caloric nor light. But the appearance of caloric and light, in many cases, shews that there is an intimate connection between them and the electric fluid. In short, they seem to form part of its composition; and there is some ground to consider it as a compound, containing caloric and light, and probably some peculiar constituent, to which we give the name of electricity.*[19]

He gave priority in this suggestion to James Russell who had been Robison's predecessor in the Chair in Natural Philosophy at Edinburgh.

---

[13] James Tytler (attr.)(n.d.), "Magnetism", *Encyclopaedia Britannica,* 3rd edn. vol.10 pp.428-448, on p.434
[14] Tytler, "Magnetism", p.433
[15] Morus, I. R. (2011). What Happened to Scientific Sensation? *European Romantic Review*, *22*(3), 389–403. https://doi.org/10.1080/10509585.2011.564463
[16] James Tytler (n.d.), "Electricity", *Encyclopaedia Britannica,* 3rd edn vol.6 pp.418-544, on p.538-9
[17] John Robison (1801), "Magnetism", *Encyclopaedia Britannica,* Supplement to the 3rd edn. vol.2 (Edinburgh) pp.112-156 on p.154
[18] Robison, "Magnetism", p.154
[19] Jeremiah Kirby (c.1805), "Electricity", *Encyclopaedia Britannica,* 4th edn. vol.7 (Edinburgh), pp.645-808 on p.757

By 1813-19 when John Playfair, Robison's successor as Professor of Natural Philosophy, wrote his over-arching "Dissertation Second On The Progress of Mathematical and Physical Science since the Revival of Letters in Europe" the possibility of establishing a connection between any of the forces of nature appeared remote.[20] Like Robison, he emphasised a mathematical approach, and he was sceptical about attempts at unification which,

> *... have occurred in every period of science. Thus, electricity has been applied to explain the motion of the heavenly bodies; and, of late, galvanism and electricity together have been held out as explaining not only the affinities of Chemistry, but the phenomena of gravitation, and the laws of vegetable and animal life.*[21]

He noted the Baconian philosophy underlying attempts at unity that proceeded through accumulation of facts and subsequent ascent from facts to laws and then grand theories, and criticised Bacon for being over-sanguine about his method:

> *An equal degree of mystery [to heat] hangs over the other properties and modifications of body; light, electricity, magnetism, elasticity, gravity, are all in the same circumstances; and the only advance that philosophy has made toward the discovery of the essences of these qualities or substances is, by exploding some theories, rather than by establishing any…. Besides this, in all the above instances the laws of action have been ascertained; the phenomena have been reduced to a few general facts, and in some cases, as in that of gravity, to one only; and for aught that yet appears,* this is the highest point which our science is destined to reach [emphasis added].[22]

Thus Playfair differed significantly from Ørsted in his views on the possible unification of forces, and he did not single electricity and magnetism out as a unique pairing as Leslie was to do a few years after Ørsted's discovery. Besides the obvious benefits of hindsight, it is possible that Leslie's was an implicitly partisan account. By the time that Leslie was writing, Ørsted's claim to priority in the discovery had been challenged (see below). As one counter to the challenge, Ørsted himself emphasised that he had suggested as long ago as 1812 that electricity and magnetism were connected, and Leslie may be upholding this claim.

Despite his possible partisanship, Leslie had fundamentally misunderstood Ørsted's work, as we can see from his account of Ørsted's experiment and his own suggested explanation.

> *The remarkable discovery of Oerstedt has greatly enlarged the field of magnetic influence. A wire of any kind of metal being laid horizontally and* at right angles to the magnetic meridian, *to connect the opposite conductors of a Galvanic Battery, a needle either below or above it is drawn considerably to the one side or the other. Instead of bewildering the imagination with* the vagaries of invisible streams, *a sufficient explication of the phaenomenon may be deduced from two leading principles:—1. Magnetism is in some proportion diffused through all metallic substances, owing either to their peculiar constitution or the universal dissemination of ferruginous molecules: 2.* The cross wire, from its position with regard to the Terrestrial Magnet, acquires induced magnetism, but extending transversely; the

---

[20] The volume containing the first half of Playfair's Dissertation was published in 1824. However, the Dissertation was cut short by Playfair's death in 1819, so the writing must predate Ørsted's discovery.
[21] John Playfair, "Dissertation Second Exhibiting a General View Of the Progress of Mathematical and Physical Science since the Revival of Letters in Europe", *Encyclopaedia Britannica*, Supplement to the 4th, 5th & 6th edns, vol.2 (Edinburgh, 1824), pp.1-127, on p.36
[22] Playfair, "Dissertation Second", pp.62-3

> under side having the virtue of a north pole, and the upper side that of a south pole. *The copious infusion of that virtue is occasioned probably by the duration of the internal tremor, excited by intense electrical action, and analogous to the effects on a bar of iron or steel subjected to hammering, twisting, beating, or the fulminating shock. Hence are easily explained the diversified phases of attraction, rotation, or impressed magnetism* [emphasis added].[23]

As in the quote from Leslie at the beginning of this section, he rejected the idea of an electric current, "the vagaries of invisible streams", and arrived at his own explanation. But, to do so, he mistook Ørsted's experimental setup, and had clearly not repeated the experiment himself. By arranging the connecting wire East-West, above a magnetised needle which is presumably pointing North-South already, he would not have seen any deflection. In his explanation, the connecting wire was magnetised by heat, like an iron bar struck by a spark. He described the "impressed magnetism" in a process that mirrored that of electrostatic induction. For Leslie, electricity was still static electricity.

To understand the concepts of Leslie and others like him, and opportunities these provided to challenge Ørsted's originality, we turn back to the *Encyclopaedias* to see what they say about current electricity.

### Contemporary understandings of current electricity

Throughout the period "galvanism" was generally used in reference to what we might think of as current electricity. Galvanism had its own separate articles in the *Encyclopaedia* and its status as some sort of "electricity" (i.e. static electricity) was contested. The word "current" was used only when invoking theories that the flow of some sort of fluid (not necessarily the electric fluid) was responsible for the observed effects.

In 1800 the article on Galvanism, probably written by John Robison, was emphatically against a premature identification of galvanism with electricity or the electric fluid:
> *Galvanism, is the name now commonly given to the influence discovered nearly eight years ago by the celebrated Galvani, professor of anatomy at Bologna, and which, by him and some other authors, has been called* animal electricity. *We prefer the former name, because we think it by no means proved, that the phenomena discovered by Galvani depend either upon the electric fluid, or upon any law of animal life. While that is the case, it is surely better to distinguish a new branch of science by the name of the inventor, than to give it an appellation which probably may, and, in our opinion, certainly does, lead to an erroneous theory* [emphasis in the original].[24]

However, James Millar in the 4th edition (c.1806) believed that the identity of the galvanic fluid and the electric fluid was sufficiently established and outlined a one fluid theory, talking of electric current in this context:
> *If a communication is established between the upper and lower plates of the pile, by means of conductors, according to the laws of electricity, the excess at the top of the pile immediately passes to the bottom. A current of electricity, therefore, will pass through the pile, and will continue….*[25]

---

[23] Leslie, "Dissertation Fourth" p629
[24] John Robison (attr.) (1800), "Galvanism", *Encyclopaedia Britannica,* Supplement to the 3rd edn. vol.1 (Edinburgh) pp.676-693 on p.676
[25] James Millar (c.1806), "Galvanism", *Encyclopaedia Britannica,* 4th edn. vol.9 (Edinburgh), pp.331-368 on p.361

Around 1816 the Frenchman, Jean Baptiste Biot, wrote the article on Galvanism for the supplement to the 4th, 5th and 6th editions. He agreed with Millar that galvanic phenomena were essentially electric, discounted the possibility of a special and different "animal electricity", but adhered to a two-fluid model that was an extension of static electrical concepts. The two "poles" of a voltaic pile were oppositely charged and he devoted much of the article to effects, such as the decomposition of water, that could be produced in a space or medium between the two poles. He just about reached a description of a closed conducting circuit, as the limit of bringing wires attached to the poles closer together. But this was not what he was interested in; he remained focused on what happened in the narrow space between the wires, and his aim was to reinforce the evidence for galvanic electricity being like that produced by static electrical machines:

> [An electric pile] *produces most remarkable effects…. If we form the communication,… between the two poles, by very fine metallic wires, and make them gently approach until they touch each other, an attraction arises between them which keeps them united in spite of the force of their elasticity. If these wires are of iron, a visible spark is excited between them, which…, produces a real combustion of the iron…. We may inflame any explosive gas, … as with the sparks from electrical machines.*[26]

He ascribed the heating and lighting effects in this space, to violent collisions between the two fluids:

> *Perhaps the electrical current, or rather, the two opposite electrical currents which meet together, and neutralise each other's effect in the substance submitted to experiment, could be conceived to act on its particles by a compression or percussion, and to extract heat in the same manner as in the boring, or flattening of metals, or in the action of the hammer.*[27]

Similar imagery was evident in Ørsted's own account of his effect. He used the term "conflict" rather than "percussion". Like Biot he sought an explanation for heat and light in this conflict. Unlike Biot, he was interested in the effects of *closed* circuits, but he was interested in opposing forces, not fluids, and did not use the term "current". For him the important point of his experiment was that it showed that the "conflict" extended outside the wire into the surrounding space, rather than being confined to the wire as might have been expected for electric fluids. His description was reminiscent of what later electricians conceptualised as a field.

> *The electric conflict can act only on the magnetic particles of matter. All nonmagnetic bodies seem to be penetrable by the electric conflict, whereas magnetic bodies, or rather their magnetic particles, seem to resist the passage of this conflict; hence, they can be moved by the impetus of the contending powers. It is sufficiently evident from the preceding observations that the electric conflict is not confined to the conductor but, as mentioned above, is dispersed quite widely in the circumjacent space.*
> 
> *I have demonstrated, in a book published seven years ago, that heat and light are an electric conflict. From the observations just reported, we may now conclude that circular motion also occurs in these effects. I believe that this will contribute very much to the elucidation of the phenomena which are called the polarization of light.*[28]

---

[26] Jean Baptiste Biot (1824), "Galvanism", *Encyclopaedia Britannica*, Supplement to the 4th, 5th & 6th edns, vol.4 (Edinburgh), pp.428-445 on p.441
[27] Biot, "Galvanism", p.442
[28] Ørsted , Experiments on the Effect of the Electric Conflict, p276

Throughout this succession of articles, from Robison to Leslie, the imagery was of violence, collision, explosion, beating and hammering. The more sensational (literally) phenomena of static electricity carried over into the concepts through which galvanism was understood. The effect of the debate over whether galvanism was a form of (static) electricity was to concentrate attention on experiments that would show that an electric pile could produce the same effects as an electrical machine, i.e., on experiments with open circuits. The fact that heat, hammering, and electric sparks could magnetise iron, while electricity could cause heat, light, and violent shocks, made it very difficult to separate out the various effects. The difficulty was compounded when light, heat, electricity and magnetism were ascribed to mysterious fluid(s) - the same fluid or different fluids according to one's philosophy.

Thus, although Ørsted's experiment caused an unprecedented scientific furore, his priority was open to challenge since the difference between what he had done and the experiments of others was not obvious to everyone. Many claimed that credit for the discovery should go to the little-known Italian Gian Domenico Romagnosi. In 1802 Romagnosi had observed deflection of a compass needle near a wire connected to one terminal of an electric pile. But it seems clear, from the detailed description of his experiment, that he was using an open circuit; like Biot and others, he was trying to produce electrostatic effects from the electric pile, proving the identity of galvanic and static electricity.[29] Similarly, in 1804 the Genoese chemist Mojon demonstrated that an electric current could magnetise a needle; this demonstrated a similarity between galvanic and static electricity, for he included the needle in his circuit so the current passed through it; the analogy was with an electric spark. The following year, in a well-publicised experiment, two Frenchmen, Jean Hachette and Bernard Desormes attempted to detect a connection between galvanism and magnetism by floating a large voltaic pile on water and seeing whether it aligned with the magnetic meridian. But they left the circuit open and the results were negative.

Ørsted, was different from any of these predecessors. He worked with a closed circuit, and his needle was outside it. He appreciated the importance of these differences and emphasised them in his initial paper:
> *It seemed demonstrated by these experiments that the magnetic needle can be moved from its position by means of a galvanic apparatus,* but by a closed galvanic circuit, not an open one*, as several very celebrated physicists tried in vain some years ago* [emphasis added].[30]

The first *Encycolpaedia* article that aligns with modern histories of Ørsted's experiment was David Brewster's "Electricity" in the 7th edition:
> *Guided by theoretical anticipations, Professor H.C. Oersted of Copenhagen, in 1820, laid the foundations of the science of* Electro-magnetism. *He found the experiment, he discovered the fundamental law, that* the magnetic effect of the Voltaic current had a circular motion round the current, *or round the* conductor, *or the wir through which the current passed. M. Ampere of Paris soon afterwards made the important discovery, that two wires conducting electrical currents...* attracted *each other when*

---
[29] Romagnosi's work is examined in detail in Martins, R. de A. (2001). "Romagnosi and Volta's Pile: Early Difficulties in the Interpretation of Voltaic Electricity". In F. Bevilacqua & L. Fregonese (Eds.), *Nuova Voltiana: Studies on Volta and his Times* (Vol. 3, pp. 81–102). http://citeseerx.ist.psu.edu/viewdoc/download?doi=10.1.1.680.8112&rep=rep1&type=pdf [accessed 8 February 2021]
[30] Ørsted, Experiments on the Effect of the Electric Conflict, p276

*the currents moved in the* same direction, *and* repelled *each other when they moved in* opposite *directions... so that, as Professor Oersted remarks,* an electric current contains a revolving action, exhibiting every appearance of polarity [emphasis in the original].[31]

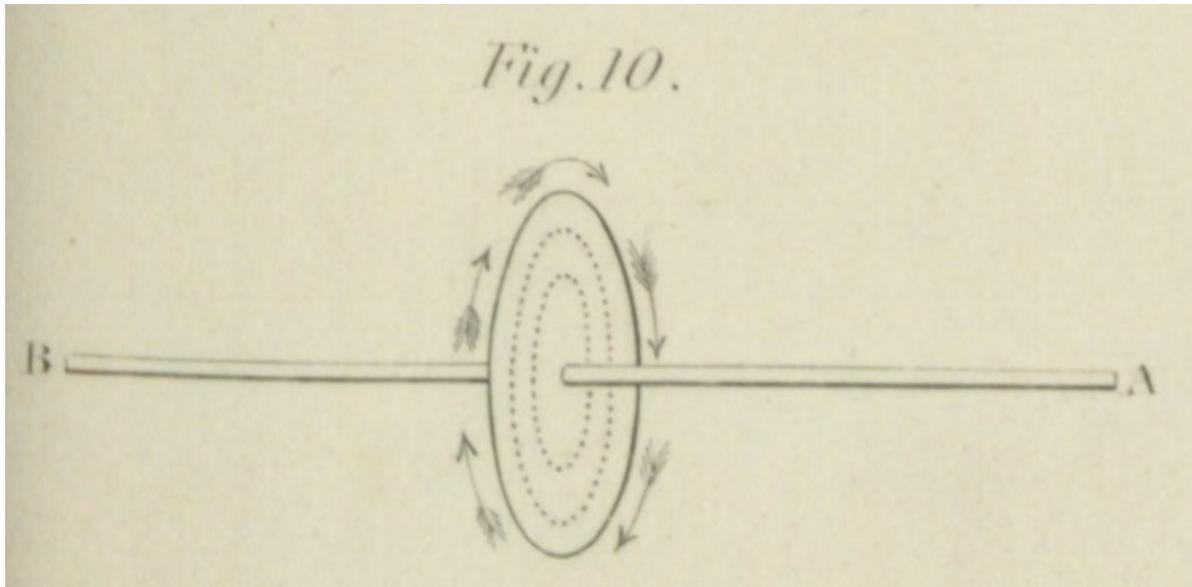

Figure 3. David Brewster's diagram of Ørsted's magnetical action. *Encyclopaedia Britannica,* 7th edn vol.8 plate CCXIII fig.10. National Library of Scotland, Creative Commons Attribution 4.0 International Licence

## Conclusion

Successive editions of the *Encyclopaedia Britannica* provide an insight into developments in contemporary thinking about electricity, magnetism, and their connection in the run-up to Ørsted's 1820 experiment and its immediate aftermath. They show that although a connection between electricity and magnetism had, indeed, been long suggested, this was far from the only candidate connection. Perhaps more interestingly, they show how difficult the development of concepts of electric currents were. They took far longer than we might imagine, with the focus generally being on experiments with open circuits designed to demonstrate the identity of galvanic and static electricity. Concepts of electric current were still not well established when Ørsted performed his experiment, and were irrelevant to Ørsted himself who preferred forces. The articles help us to appreciate ways in which misinterpretation of these experiments laid his priority open to challenge.

---

[31] David Brewster (1842), "Electricity", *Encyclopaedia Britannica,* 7th edn vol.8 (Edinburgh), pp.565-663 on p.574